\documentclass[conference]{IEEEtran}
\IEEEoverridecommandlockouts
\usepackage{amsmath,amssymb,amsfonts}
\usepackage{algorithmic}
\usepackage{graphicx}
\usepackage[inline]{enumitem}

\usepackage{textcomp}
\usepackage{xcolor}

\usepackage{lipsum}
\usepackage{multicol}
\usepackage{soul}
\usepackage{hyperref}
\usepackage[disable]{todonotes}
\usepackage[sort&compress,numbers]{natbib}

\usepackage{setspace}
\usepackage{comment}

\usepackage[ruled,vlined,linesnumbered]{algorithm2e}

\usepackage{subfig}
\usepackage{array}
\usepackage{longtable}

\usepackage{booktabs}

\def\BibTeX{{\rm B\kern-.05em{\sc i\kern-.025em b}\kern-.08em
    T\kern-.1667em\lower.7ex\hbox{E}\kern-.125emX}}

\usepackage{fancyhdr} % see https://tex.stackexchange.com/questions/64667/text-in-footer-on-first-page-but-no-page-numbers
\begin{document}

% See references.bib to control maximum number of authors: https://tex.stackexchange.com/questions/164017/limiting-the-number-of-authors-in-the-references-with-ieeetran?noredirect=1&lq=1
\bstctlcite{IEEEexample:BSTcontrol}

\title{Managing Data Lineage of O\&G Machine Learning Models: The Sweet Spot for Shale Use Case}

\author{
\IEEEauthorblockN{
    Raphael Melo Thiago,
    Renan Souza,
    Leonardo Azevedo,
    Eltons Soares,
    Rodrigo Santos,
    Wallas dos Santos,
}
\IEEEauthorblockN{
    Maximilien de Bayser,
    Marcelo Costalonga Cardoso,
    Marcio F. Moreno,
    Renato Cerqueira
}
\IEEEauthorblockA
{
IBM Research
}
}

\maketitle

%%% MY COMMANDS

\newcommand{\needref}[1][?]{\colorbox{lightgray}{[R #1]}}
% Usage: \needref{} or \needref[any hint on the refs]{}

\newcommand{\eg}{\emph{e.g.},}
\newcommand{\egUpper}{\emph{E.g.},}
\newcommand{\ie}{\emph{i.e.},}
\newcommand{\etal}{\textit{et al.}}

\newcommand{\codebackground}[1]{\colorbox{black!5}{\parbox{\dimexpr\linewidth-2\fboxsep}{\fontfamily{pcr}\scriptsize#1}}}

\newcommand{\codefont}[1]{{\fontfamily{pcr}{\small{#1}}}}

% Counter
\newcounter{qcounter} 
\newcommand{\createQ}[2]{
    \refstepcounter{qcounter} \label{#1} \textit{\textbf{Q\ref{#1}:} {#2}}
}
\newcommand{\refQ}[1]{Q#1}
% Usage: first create the reference using \createQ{label_name}, then you can make a reference to it using refQ{label_name}

\newcommand{\MLCycle}{ML lifecycle in CSE}
\newcommand{\queries}{\hyperref[tab:queries]{Q1--Q6}}
\newcommand{\textonto}[1]{\texttt{\small{#1}}}
\newcommand{\textontohead}[1]{\textbf{\emph{#1}}}
\newcommand{\textsoftware}[1]{\textsc{#1}}

\newcommand{\alert}[1]{\textcolor{red}{#1}}

\newcommand{\provcapturesystems}{\cite{komadu_suriarachchi_big_2018, lucas_carvalho2018provcompliant, joao_survey_2019, silva_raw_2017, silva_capturing_2018}}

\begin{abstract}
Machine Learning (ML) has increased its role, becoming essential in several industries. However, questions around training data lineage, such as "where has the dataset used to train this model come from?"; the introduction of several new data protection legislation; and, the need for data governance requirements, have hindered the adoption of ML models in the real world. 
In this paper, we discuss how data lineage can be leveraged to benefit the ML lifecycle to build ML models to discover sweet-spots for shale oil and gas production, a major application in the Oil and Gas (O\&G) Industry.
\end{abstract}


\thispagestyle{fancy}%

\section{Introduction}
In this paper, we discuss how data lineage can be leveraged to benefit the Machine Learning (ML) lifecycle to build ML models to discover sweet-spots for shale oil and gas production, a major application for the Oil and Gas (O\&G) Industry.

ML is an artificial intelligence discipline that aims at producing predictive models to perform tasks without the need to explicitly defined them~\cite{bishop2006pattern}. Models are learned, or generalized, from a set of training data, this data is usually derived from raw or closely related to the task's domain, e.g., a house price prediction model would need to be trained using datasets derived from real house price data\footnote{Like the Boston house-prices dataset.}, passed through one or a series of data transformations.

However, to obtain a reliable ML model, one usually needs to perform several iterations of the ML lifecycle~\cite{prov2019works}  -- obtaining relevant raw data, transforming data, training and validating an ML model. Given the heterogeneous nature of this lifecycle, it is difficult to track how data is transformed through it. In practice, tracking data transformations is usually done manually, which is an error-prone and
time-consuming task. The increasing use of ML models in several industries, coupled with the introduction of several regulatory pieces concerning data privacy and protection, like the European's General
Data Protection Regulation (GDPR), the California Consumer Privacy Act (CCPA), and Brazil's Lei Geral de Prote\c{c}\~ao de Dados (LGPD), among others, as well as industry-specific data concerns, like the ones faced in the O\&G domain,
makes the problem of having reliable track of data transformations one of the top concerns for a model's adoption.

Data lineage, or data provenance, helps reproducing, tracing, assessing, and understanding data and their transformation processes~\cite{herschel2017survey}. In this paper, we present how a provenance data tracking and analysis system, called ProvLake \cite{souza2019efficient}, can aid building models for finding Sweet Spot for Shale (SSS) \cite{bianca2019sss}. ProvLake ~\cite{souza2019efficient} supports inter-workflow tracking of heterogeneous data stored in heterogeneous stores with design principles suited for the tracking data in ML lifecycles \cite{prov2019works,souza_managing_2019}.

This paper starts with the current introduction, followed by a description of ProvLake. Then we introduce and explain the Sweet Spot for Shale use case and how ProvLake can be used within it. Afterwards, it is presented some
experimental analysis and, finally, our concluding remarks.

\section{ProvLake}
ProvLake~\cite{souza2019efficient} is a provenance tracking and analysis system for runtime analysis of multi workflow data. It was built using a set of design principles that enables it to perform well in the
ML lifecycle: (i)~\textit{asynchronicity} and the use of (ii)~\textit{work queues} allow it to have a small (time-wise) provenance capture overhead and its (iii)~\textit{lightweight library} makes so the original code remains as close as possible to its original form.

\begin{figure}[!htb]
  \centering
  \includegraphics[width=0.5\textwidth]{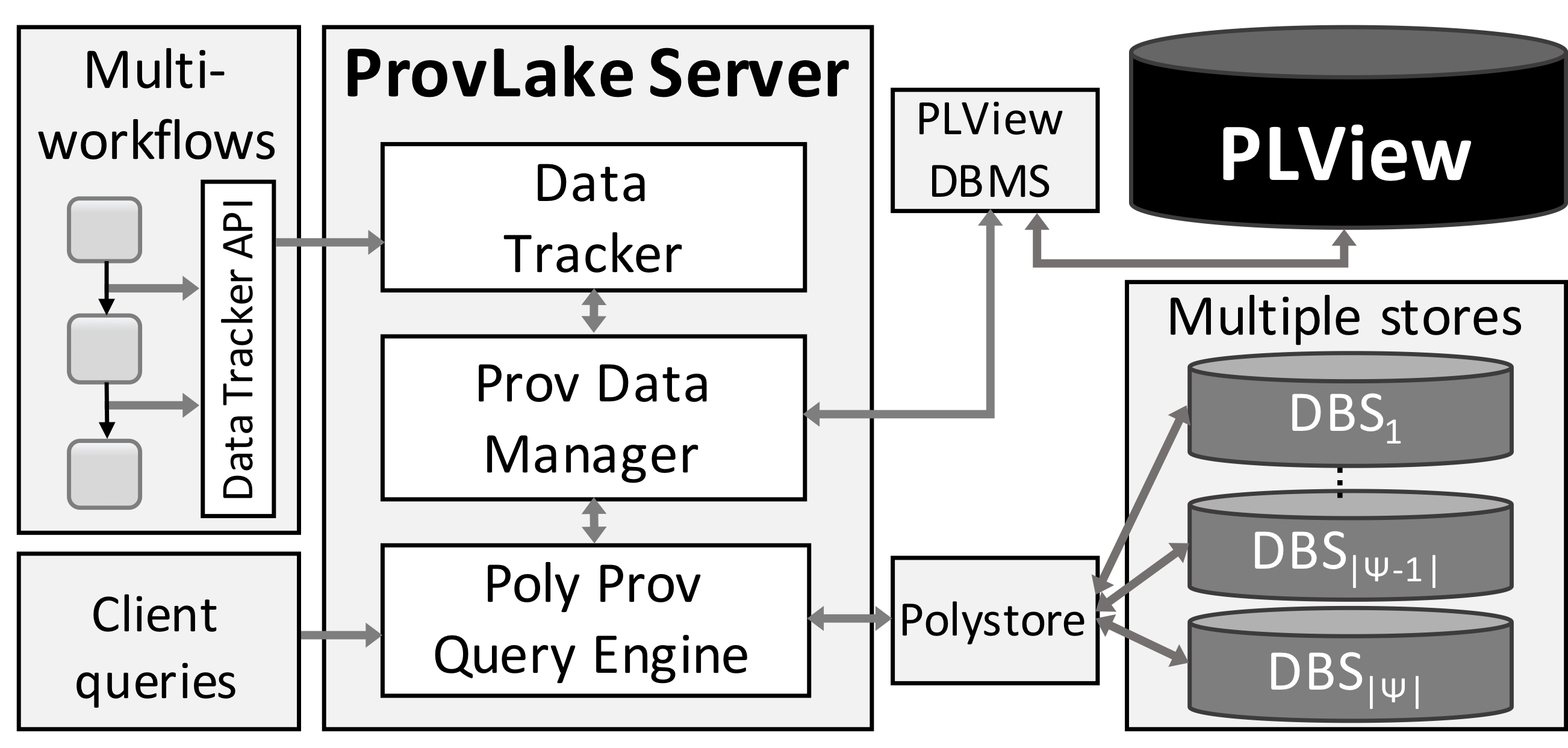}
  \caption{ProvLake Architecture \cite{souza2019efficient}.}\label{img:provlake:architecture}
\end{figure}

ProvLake has a microservices architecture composed of three services (Data Tracker, ProvManager, and PolyProvQueryEngine), a lightweight Data Tracker API, a messaging system, and the ProvLake Data View (PLView) (Figure~\ref{img:provlake:architecture}). These components capture data, transform them into the provenance database representation, insert them into the database, and help runtime data analysis through query submissions.

ProvLake logically integrates and ingests multiworkflow data into the PLView ~\cite{souza2019efficient}. PLView holds domain data extracted from data stored in multiple stores; explicit
data relationships between datasets distributed across multiple stores; and the multi workflow data relationships. PLView is materialized in a DBMS. Data is stored according to a data model that extends W3C's PROV specification \cite{provdm} with ML-specific terms and concerns. 
Retrospective provenance data, that is the data captured during workflow execution \cite{herschel2017survey}, can be interpreted as a directed graph, where nodes are either data items or data transformations (activities) and edges are data dependencies. Therefore, asking provenance queries is the same as finding a path between two vertices in a graph. Table~\ref{tab:queries} presents some examples of provenance queries in the O\&G domain.

\begin{table*}
\caption{Example of provenance queries for SSS' use case}
\label{tab:queries}
\resizebox{\textwidth}{!}{%
\begin{tabular}{ll}
Q1 & Which trained model depends on the production data of a particular well?                                     \\
Q2 & Which trained model depends on a given well?                                                                 \\
Q3 & For a given zone, what is the best trained model, i.e., the model with the smallest mean square error (MSE)?
\end{tabular}%
}
\end{table*}

\section {The sweet spot for shale use case}

The sweet-spot for shale identification task aims at finding one or a set of drilling locations within a play that has a high production potential.

Guevara et al. \cite{bianca2019sss} present a machine learning methodology to integrate well-data for sweet-spot identification. The methodology integrates data from different data sources: petrophysical data from horizontal- and vertical-well logs, completions data, engineering data, and production data from horizontal wells. From this, it generates a map from petrophysical data using ML feature extraction techniques to some specific geological zone, also producing an integrated data set that will contain completion, engineering, and petrophysical-feature as feature values and the cumulative production of oil and gas within the producing months of interest as target values. This integrated data set is then fed to a predictive modeling engine -- where this data is then used to create different kinds of machine learning models (black-box models, domain-knowledge constrained models, and models to construct hypothetical \textit{what-if} scenarios). Lastly, the outputs of trained models are used for performing effect analysis of the covariates on the response variable for sweet-spot identification.

\section{Provenance in the Sweet Spot for Shale use Case}

\begin{figure*}
  \centering
  \includegraphics[width=\textwidth,keepaspectratio]{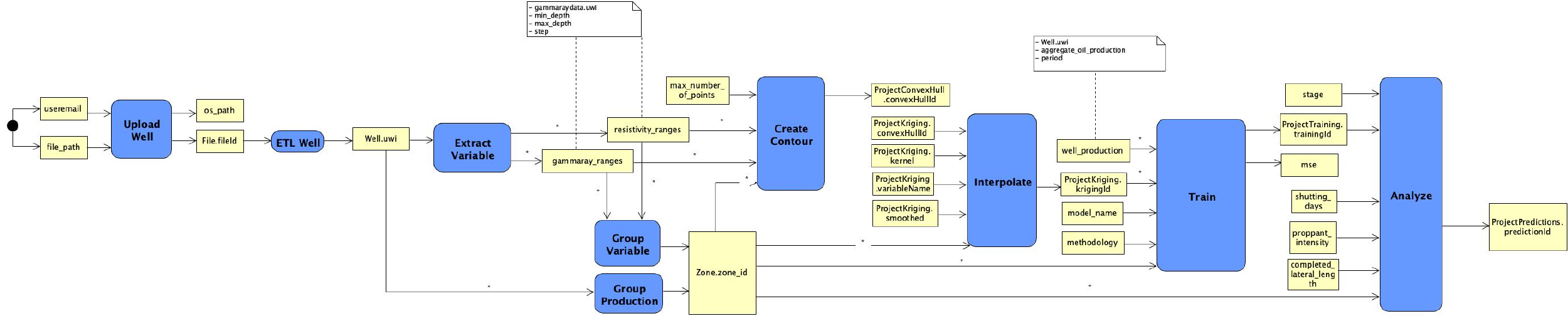}
  \caption{Sweet Spot for Shale Workflow. Yellow boxes are data items, rounded blue boxes are data transformations, and arrows represent data dependencies -- an arrow leaving a data item indicates that the following activity used it, while an arrow arriving indicates that the activity generated it.}\label{img:sss:workflow}
\end{figure*}

We integrated ProvLake into a system that implemented the previous machine learning methodology as a series of microservices, henceforth referred to as SSS. Following the multidisciplinary team methodology put forward by \cite{souza2019efficient}, we modeled the prospective provenance for the system, which workflow is presented in  Figure~\ref{img:sss:workflow}. Prospective provenance is the contextual data, \textit{i.e.}, data that does not change between workflow executions \cite{herschel2017survey}, in ProvLake, the prospective provenance is used to describe the workflow. To aid in this process, we gathered a set of key relevant queries (Table~\ref{tab:queries} depicts a subset of them).

% An example of query answer given the "prospective provenance"

Each of SSS' microservices could be mapped to one workflow. Because of its architecture and a stable API, instead of instrumenting each microservice separately, we instead instrumented the middleware responsible for handling requests. Thus, decoupling instrumentation code from the actual code performing the machine-learning tasks.

Aside from the online retrospective provenance tracking, we also derived retrospective provenance from the historical data present in SSS' records. This was only possible because stored data schema had the necessary information to generate provenance data (following the prospective provenance). 

\begin{figure*}
  \centering
  \includegraphics[height=4.5cm,keepaspectratio]{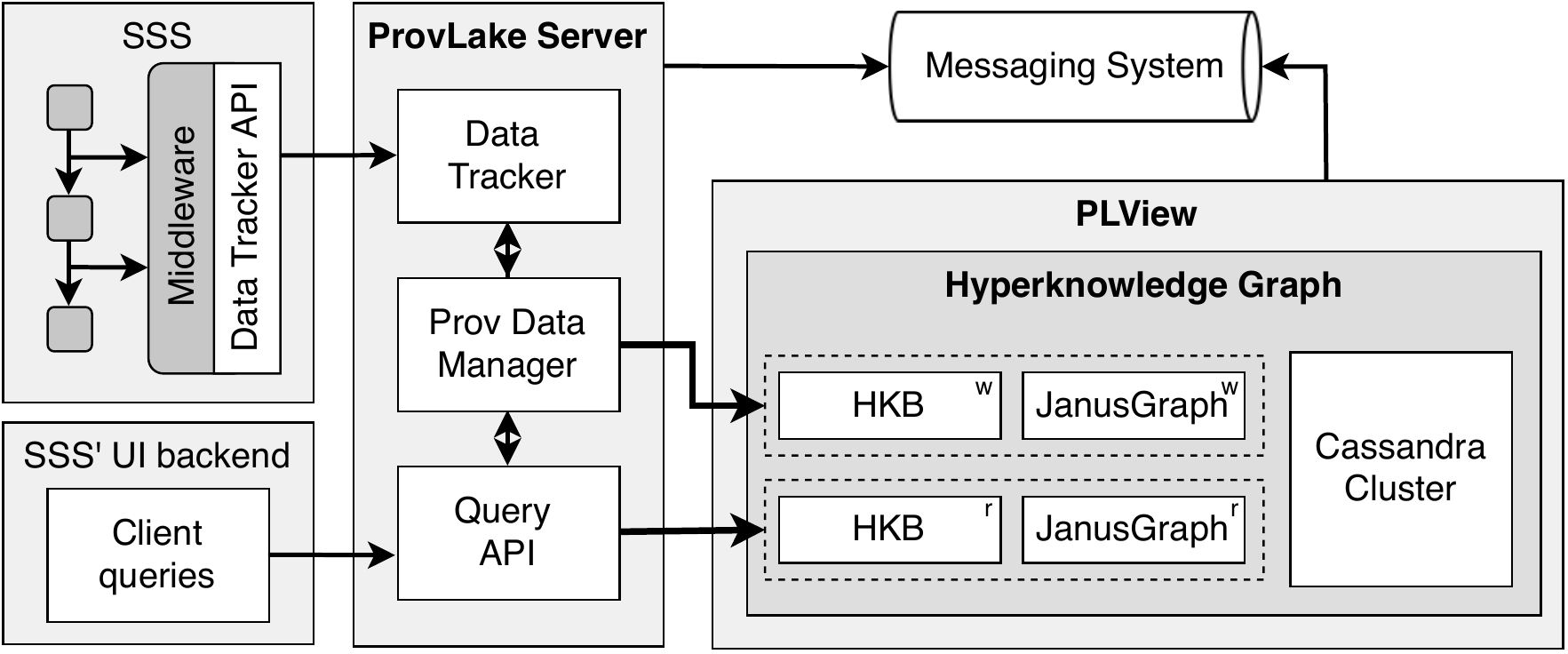}
  \caption{ProvLake's architecture for the SSS use case. A HyperKnowledge Base (HKB)~\cite{7823600} instantiates the PLView. A Messaging system was used for tracking the progress of asynchronous insertions. In this particular use case, the ProvLakeLib was used to instrument SSS' middleware instead of its scripts.}\label{img:provlake:actual}
\end{figure*}

However, ingesting historical data into ProvLake posed a challenge, the insertion profile differs substantially from the normal provenance tracking -- data was ingested in a batch like fashion, at fixed intervals we ingested a constant number of records. This batch ingestion process was sufficiently high to stress the Database Management System (DBMS) used to store provenance data (in this implementation, we employed JanusGraph\footnote{https://janusgraph.org/} DBMS) -- even though insertions were done asynchronously, they were affecting the, necessarily synchronous, queries. Because of this requirement, we employed a Command-Query Responsibility Segregation (CQRS)~\cite{newman2015building} like approach in the access of the provenance database. Isolating information retrieval (query-side) from transaction (command-side) operations by having one database service for the first and another one for the second; however, we kept both servers accessing the shared database. Besides, the database was replicated. In other words, considering our technology stack, our deployment included one JanusGraph for reading operations, one JanusGraph for writing operations, and three Cassandras\footnote{http://cassandra.apache.org/} as database replicas, Figure~\ref{img:provlake:actual} depicts the instantiated architecture.

% \section{ProvLake and SSS or the Experimental justification for the architecture}
% \input{sections/prov_sss_experiments.tex}

A general query API was designed to answer most provenance queries. To query, one defines an identified starting vertex, or \textit{seed}, and a destination vertex, or \textit{target}. Both \textit{seed} and \textit{target} vertices can be type restricted, where types can either be from ProvLake's ontology or the domain (prospective provenance). Targets do not need to be identified. Users can also choose the traversal direction (\textit{forward} or \textit{backward}). Figure~\ref{img:query:api} shows examples of how the query API can be used to answer provenance queries.

% \begin{figure}[!htb]
%   \centering
%   \includegraphics[width=1\textwidth]{images/query_api.pdf}
%   \caption{Example of query requests answering queries in Table~\ref{tab:queries}. The query API follows the REST principle, where each subpath is subordinate to the previous, e.g., in Q1, one will find a provenance seed of the type \texttt{WELL} with \texttt{values} (identifier) "well1232" traversing the graph \texttt{forward} until to find targets of the type \texttt{PROJECTTRAINING}. }\label{img:query:api}
% \end{figure}

\begin{figure*}
  \centering
% Please add the following required packages to your document preamble:
% \usepackage{graphicx}
% \begin{table}[]
% \caption{Example of provenance queries for SSS' use case}
% \label{tab:queries}
\resizebox{\textwidth}{!}{%
\begin{tabular}{ll}
Q1             & Which trained model depends on the production data of a particular well?                                                \\
\multicolumn{2}{l}{
    \texttt{/provenance/seeds/WELL/values/12153/direction/forward/targets/PROJECTTRAINING}
} \\
  &                                                                                                                         \\
Q2             & Which trained model depends on a given well?                                                                            \\
\multicolumn{2}{l}{
    \texttt{/provenance/seeds/os\_path/values/file\_158.las/direction/forward/targets/PROJECTTRAINING}
} \\
  &                                                                                                                         \\
Q3             & For a given zone, what is the best trained model, i.e., the model with the smallest mean square error (MSE)?            \\
\multicolumn{2}{l}{
    \texttt{/provenance/seeds/ZONE/values/278/direction/forward/targets/PROJECTTRAINING}
}
\end{tabular}%
}
% \end{table}
 
  \caption{Example of query requests answering queries in Table~\ref{tab:queries}. The query API follows the REST principle, where each subpath is subordinate to the previous, e.g., in Q1, one will find a provenance seed of the type \texttt{WELL} with \texttt{values} (identifier) "12153" traversing the graph \texttt{forward} until to find targets of the type \texttt{PROJECTTRAINING}. }\label{img:query:api}
\end{figure*}

% \section{Analyzing SSS provenance data}
% \input{sections/analyzing_sss_prov.tex}

\section{Conclusions}
% This is the first sentence of the conclusions.

% \begin{enumerate}
%     \item Recap da motiva\c{c}\~ao
%     \begin{enumerate}
%         \item Como o artigo lidou com ela?
%     \end{enumerate}
%     \item Listar principais contribui\c{c}\~oes
%     \begin{enumerate}
%         \item Gerais 
%         \item para O\&G
%     \end{enumerate}
%     \item Trabalhos futuros*
% \end{enumerate}

The increasing concerns surrounding data lineage impact the machine learning lifecycle directly. In this work, we presented the provenance tracking for a machine learning task in the oil and gas industry, the sweet spot for shale identification task. In it, a machine learning methodology is employed to predict high production potential drilling locations. We showed how ProvLake's, a provenance tracking system, design principles enabled it to cope with the sweet spot for shale system and provenance requirements.

We presented a schematic view of the sweet spot for shale prospective provenance, generated using a multidisciplinary team methodology, a process driven by knowledge from domain experts. Also, how SSS provenance characteristics affected the actual architectural implementation of the system.

Lastly, a general query API for provenance data was put forward. Given provenance data's graph-like features, in this query API to find query responses is to find a traversal path within a graph, 

\vspace{-1.0mm}
\bibliography{references}
\bibliographystyle{IEEEtran.bst}

\end{document}